\documentclass[pdftex,twocolumn,epjc3]{svjour3}          
\RequirePackage[T1]{fontenc}

\smartqed  

\RequirePackage{graphicx}
\RequirePackage{mathptmx}      
\RequirePackage{flushend}
\RequirePackage[numbers,sort&compress]{natbib}
\RequirePackage[colorlinks,citecolor=blue,urlcolor=blue,linkcolor=blue]{hyperref}

\journalname{Eur. Phys. J. C}

\begin{document}

\title{Long term monitoring of the optical background in the Capo Passero deep-sea site with the NEMO tower prototype}


\author{
	S.~Adri\'{a}n-Mart\'{i}nez\thanksref{UPV}
	\and
	S.~Aiello\thanksref{INFNCT}
	\and
	F.~Ameli\thanksref{INFNRM}
	\and
	M.~Anghinolfi\thanksref{INFNGE}
	\and
	M.~Ardid\thanksref{UPV}
	\and
	G.~Barbarino\thanksref{INFNNA,UNINA}
	\and
	E.~Barbarito\thanksref{INFNBA,UNIBA}
	\and
	F.C.T.~Barbato\thanksref{INFNNA,UNINA}
	\and
	N.~Beverini\thanksref{INFNPI,UNIPI}
	\and
	S.~Biagi\thanksref{INFNLNS}
	\and
	A.~Biagioni\thanksref{INFNRM}
	\and
	B.~Bouhadef\thanksref{INFNPI}
	\and
	C.~Bozza\thanksref{INFNSA,UNISA}
	\and
	G.~Cacopardo\thanksref{INFNLNS}
	\and
	M.~Calamai\thanksref{INFNPI,UNIPI}
	\and
	C.~Cal\`{\i}\thanksref{INFNLNS}
	\and
	D.~Calvo\thanksref{IFIC}
	\and
	A.~Capone\thanksref{INFNRM,UNIRM}
	\and
	F.~Caruso\thanksref{INFNLNS}
	\and
	A.~Ceres\thanksref{INFNBA}
	\and
	T.~Chiarusi\thanksref{INFNBO}
	\and
	M.~Circella\thanksref{INFNBA}
	\and
	R.~Cocimano\thanksref{INFNLNS}
	\and
	R.~Coniglione\thanksref{INFNLNS}
	\and
	M.~Costa\thanksref{INFNLNS}
	\and
	G.~Cuttone\thanksref{INFNLNS}
% ...
}

\thankstext{corr1}{e-mail: pellegriti@lns.infn.it}
\thankstext{corr2}{e-mail: piattelli@lns.infn.it}
\thankstext{presentaddress1}{Present address: Nikhef, Science Park, Amsterdam, The Netherlands}

\institute{INFN Sezione Bari, Via E. Orabona 4, 70126, Bari, Italy \label{INFNBA}
        \and
        INFN Sezione Bologna, V.le Berti Pichat 6/2, 40127, Bologna, Italy \label{INFNBO}
        \and
        INFN Laboratori Nazionali del Sud, Via S.Sofia 62, 95123, Catania, Italy \label{INFNLNS}
        \and
        INFN Sezione Catania, Via S. Sofia 64, 95123, Catania, Italy \label{INFNCT}
% ...
\and
        INFN Sezione Genova, Via Dodecaneso 33, 16146, Genova, Italy \label{INFNGE}
        \and
        IINFN Sezione Napoli, Via Cintia, 80126, Napoli, Italy \label{INFNNA}
        \and
        INFN Sezione Pisa, Polo Fibonacci, Largo Bruno Pontecorvo 3, 56127, Pisa, Italy \label{INFNPI}
        \and
        INFN Sezione Roma, P.le A. Moro 2, 00185, Roma, Italy \label{INFNRM}
        \and
        INFN Gruppo Collegato di Salerno, Via Giovanni Paolo II 132, 84084 Fisciano, Italy \label{INFNSA}
        \and
        Dipartimento Interateneo di Fisica Universit\`a di Bari, Via E. Orabona 4, 70126, Bari, Italy \label{UNIBA}
% ...
\and
        Dipartimento di Scienze Fisiche Universit\`a di Napoli, Via Cintia, 80126, Napoli, Italy \label{UNINA}
        \and
        Dipartimento di Fisica Universit\`a di Pisa, Polo Fibonacci, Largo Bruno Pontecorvo 3, 56127, Pisa, Italy \label{UNIPI}
        \and
        Dipartimento di Fisica Universit\`a ``Sapienza'', P.le A. Moro 2, 00185, Roma, Italy \label{UNIRM}
        \and
        Dipartimento di Fisica Universit\`a di Salerno, Via Giovanni Paolo II 132, 84084 Fisciano, Italy  \label{UNISA}
% ...
\and
        Instituto de Investigaci\'{o}n para la Gesti\'{o}n Integrada de las Zonas Costeras, Universitat Polit\`{e}cnica de Val\`{e}ncia,~Gandia,~Spain \label{UPV}
        \and
        IFIC-Instituto de F\'{i}sica Corpuscular,~(CSIC-Universitat de Val\`{e}ncia), Val\`{e}ncia, Spain \label{IFIC}
% ...
}

\date{Received:  / Accepted:}

\maketitle

\begin{abstract}
 
The NEMO Phase-2 tower 
is the first detector which was operated underwater for
more than one year at the ``record'' depth of 3500 m.
It was designed and built within the framework of the NEMO (NEutrino Mediterranean Observatory) project.
The 380 m high tower was successfully installed in March 2013 80 km offshore Capo Passero (Italy).
This is the first prototype operated on the site where the italian node of the KM3NeT neutrino telescope will be built.
The installation and operation of the NEMO Phase-2 tower has proven the functionality of the
infrastructure and the operability at 3500 m depth.
A more than one year long monitoring of the deep water characteristics of the site has been also provided.
In this paper the infrastructure and the tower structure and instrumentation are described.
The results of long term optical background measurements are presented.
The rates show stable and low baseline values, compatible with the contribution 
of $^{40}$K light emission, with a small percentage of light bursts due to bioluminescence. All these features 
confirm the stability and good optical properties of the site.

\end{abstract}

\section{Introduction}

High energy neutrinos are expected to be messengers from astrophysical objects where hadronic interactions may take place. In this framework several projects were developed (see refs. \cite{ageron2011,aynutdnov2011,achterberg2006}).
The discovery of a cosmic neutrino flux reported by IceCube initiated the era of Neutrino Astronomy. This result, at first based on an analysis performed selecting cascade contained events or track events starting in the detector (the so called HESE analysis) \cite{aartsen2013,aartsen2014}, was confirmed by the observation of a neutrino cosmic flux of upgoing muon neutrinos \cite{aartsen2015}. These fluxes are obtained with independent analyses in two different sets of events and show rather close values when normalized to one flavor.       
However, most of the questions raised about the origin of the observed neutrino cosmic flux remain unsolved. Many hypotheses have been proposed including galactic and extragalactic sources, Dark Matter etc.
As a consequence of this discovery, the physics case for the construction of a km$^3$-scale neutrino telescope in the Northern Hemisphere is strongly enforced.
A telescope in the Mediterranean Sea has a field of view for up-going neutrinos, that are the golden channel for neutrino astronomy, of 87$\%$ thus covering a very large part of the of the Galactic Plane including the Galactic Centre. 
Observations put in evidence the presence or several gamma-ray sources in the GeV-TeV energy range nearby the Galactic Centre. Since the origin of these emissions is unknown the observation of neutrinos from these regions can have a unique role in clarifying the production mechanisms. Moreover, the very good angular resolution achievable thanks to the large effective scattering length in deep sea water is one of the most distinct feature of a cubic kilometer telescope installed in the Mediterranean Sea and it is very important for the individuation of sources. 

Since 1998, the NEMO (NEutrino Mediterranean Observatory) collaboration carried out research activities 
aimed at developing and validating key technologies for a km$^3$-scale underwater neutrino telescope \cite{migneco2006,migneco2008,capone2009,taiuti2011} 
as well as searching and characterizing a deep-sea site suitable for the installation of the detector. 
The first pilot project, NEMO Phase-1, dates back to 2007 when a prototype tower was deployed at a depth of 2000 m in a Test Site
located about 20 km offshore Catania (Italy). 
Technical details and the scientific results can be found in \cite{aiello2010}.

\begin{figure}
\begin{minipage}{\columnwidth}
\includegraphics[width=1.0\textwidth]{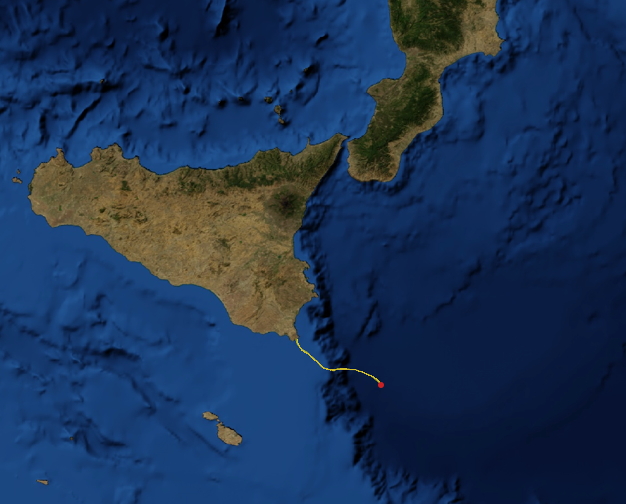}%
\end{minipage}
\caption{Map of the western Ionian region showing the location of the KM3NeT-It deep-sea site and the
electro-optical cable route.}
\label{sitemap}
\end{figure}

A deep-sea site at a depth of 3500 m, about 80 km offshore  Portopalo di Capo Passero
(Italy) ($36^\circ~17'~48''$~N, $15^\circ~58'~42''$~E), was identified as optimal for the installation of the underwater neutrino telescope (Fig~\ref{sitemap}). In the following it is indicated as the KM3NeT-It site.

In 2008 a 100 km electro-optical cable connecting the KM3NeT-It site to an on-shore station, located in Portopalo di Capo Passero, was deployed.
In the following years an important R\&D activity aimed at validating new technologies was carried out thanks to this infrastructure.
These activities culminated in the construction, deployment and operation of the
NEMO Phase-2 tower in March 2013.
The Phase-2 tower continuously took
data until August 2014 when it was disconnected to allow for an upgrade of the underwater infrastructure in preparation
for the installation of the first group of detection structures that will form the italian node of the KM3NeT detector.

The paper is structured as follows.
In Section 2 the Capo Passero site and infrastructure are described. Section 3 describes the detection tower and the details of all the included sub-systems.
Section 4 describes the installation and the sea operations, which are necessary for the long-term measurements of the optical background presented in Section 5. Conclusion and perspectives are given in Section 6.

\section{The KM3NeT-It site and infrastructure}
\label{infrastructure}

The KM3NeT-It site was characterised with
a series of sea campaigns to study and monitor the most important environmental parameters.
Oceanographical properties of the site, like deep-sea water optical properties (absorption and diffusion), water environmental properties (temperature, salinity), biological activity, optical background, water currents, sedimentation and seabed nature were studied \cite{capone2002,riccobene2007,rubino2012}. 
This activity has confirmed that the KM3NeT-It  site has optimal characteristics to host the italian node of the KM3NeT research infrastructure \cite{km3netweb}. 

The KM3NeT-It infrastructure includes a shore station, a 100 km long main electro-optical cable (MEOC) and a deep-sea installation comprising a novel design 10 kW DC voltage converter.
The onshore infrastructure, located in Portopalo di Capo Passero, hosts the power feeding system, which directly feeds the MEOC \cite{sedita2006}, as well as the control centre and the data acquisition system.

\begin{figure}
\begin{minipage}{\columnwidth}
\includegraphics[width=1.\textwidth]{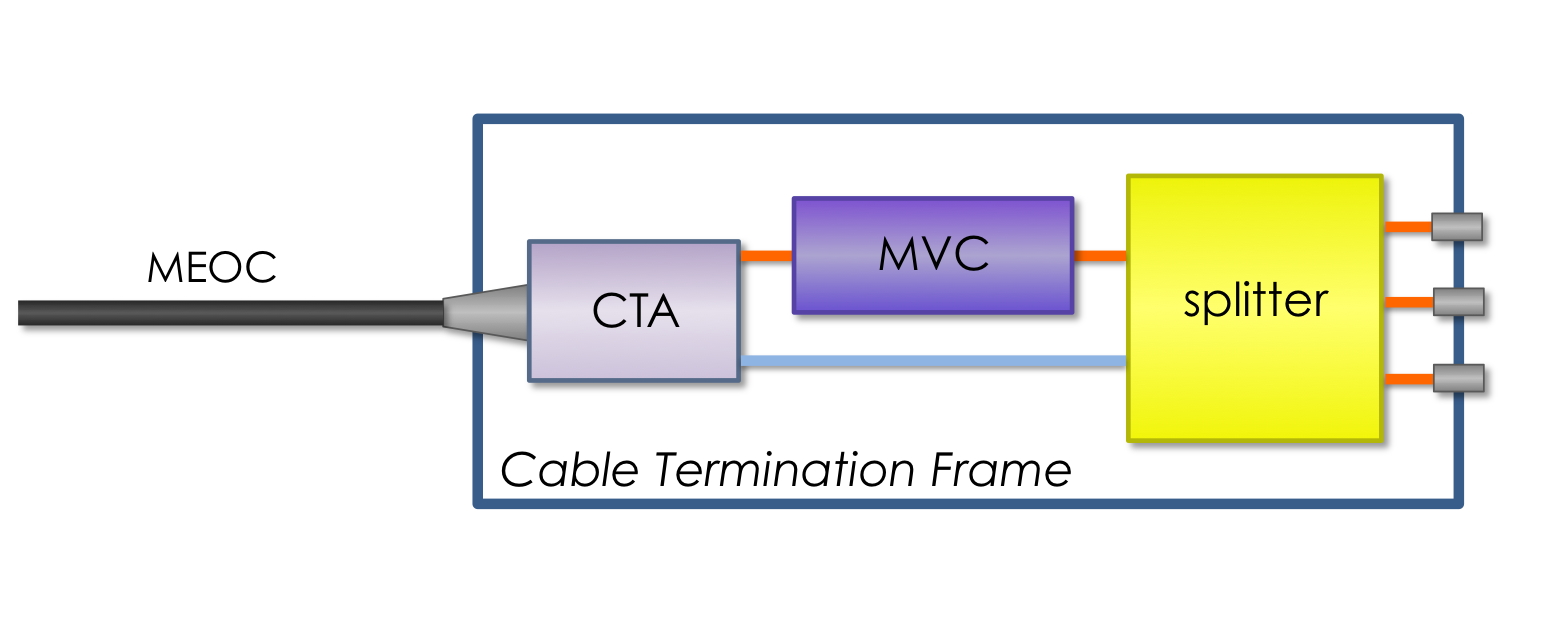}%
\end{minipage}
\caption{Scheme of the Cable Termination Frame. Power lines and optical fibres of the Main Electro Optical Cable (MEOC) are split by a Cable Termination Assembly (CTA). The 10 kW DC/DC Medium Voltage Converter (MVC) steps down the voltage from 10 kV to 375 V. A splitter assembly reroutes power lines and optical fibres towards three wet-mateable electro-optical connectors.}
\label{ctfscheme}
\end{figure}

\begin{figure}
\begin{minipage}{\columnwidth}
\includegraphics[width=1.\textwidth]{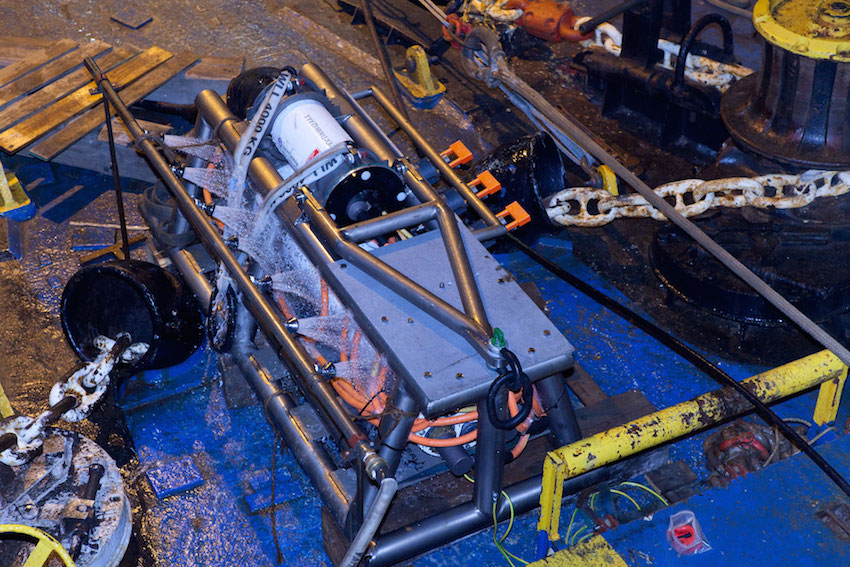}%
\end{minipage}
\caption{An image of Cable Termination Frame on the ship deck prior to its deployment.}
\label{ctfphoto}
\end{figure}

The MEOC is a standard submarine communication cable manufactured by Alcatel\footnote{Model OALC-4 17 mm Type 30.} containing 20 G655 optical
fibres\footnote{Corning Vascade LEAF negative dispersion NZ-DSF.}. These fibres have a maximum attenuation of 0.23 dB/km at 1550 nm wavelenght.
The cable has a single conductor for power transmission. Current return is provided via seawater, which offers an extremely small resistance thus reducing power losses \cite{cocimano2009}. The system incorporates sea electrodes both at the shore (anode) and at the deep-sea (cathode) ends.
At the deep-sea end the MEOC is terminated by a Cable Termination Frame (CTF) (Figs.~ \ref{ctfscheme},\ref{ctfphoto}) hosting a Cable Termination Assembly (CTA), allowing to split and reroute the fibres towards a connection panel, and a DC Medium Voltage Converter (MVC) to step down the 10 kV provided by the onshore Power Feeding System to 375 V \cite{orlando2009}.
The intermediate voltage is distributed to three output ports equipped with wet-mateable connectors.
Similarly optical fibre communication is distributed to the same output ports.
Connections are performed by means of a Remotely Operated Vehicle (ROV).

In the configuration in place at the time when the prototype described in this paper was operated, only eight out of the twenty optical fibres of the MEOC were unfolded and distributed to the three outputs available on the CTF.
In July 2015 a new CTF that uses an MVC identical to the previous one has been successfully installed.
This new CTF includes a splitter able to distribute all the twenty available fibers of the MEOC to five electro-optical output and
will be used as main distribution node of the KM3NeT-It phase-one detector.

\section{The NEMO Phase-2 tower}

\begin{figure} 
\begin{minipage}{\columnwidth}
\includegraphics[width=1.\textwidth]{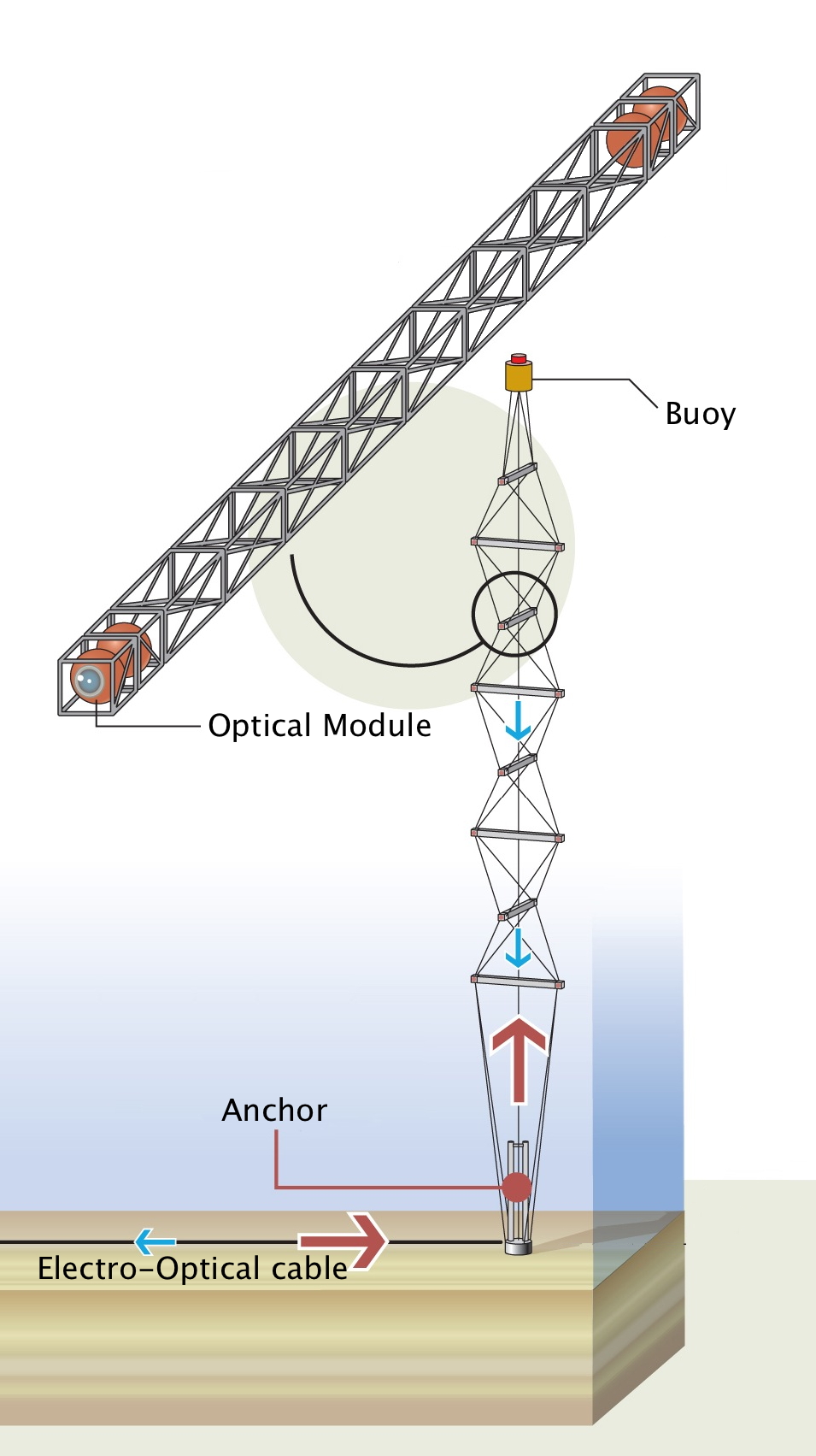}%
\end{minipage}
\caption{Sketch of the NEMO Phase-2 tower. The structure is formed by a sequence of eight floors, each one supporting four optical modules.
The structure is anchored at the seabed and kept taut by a buoy located at the top. Vertical distances are not to scale.}
\label{towersk}
\end{figure}

\begin{figure} 
\begin{minipage}{\columnwidth}
\includegraphics[width=1.\textwidth]{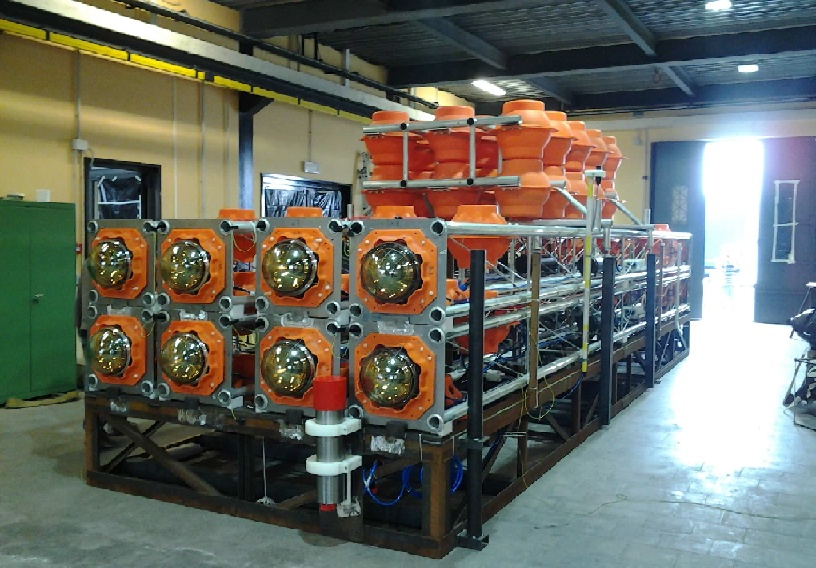}%
\end{minipage}
\caption{The NEMO Phase-2 tower in its compact configuration during the integration phase.}
\label{torre}
\end{figure}

The detection unit deployed in the KM3NeT-It site at a depth of 3449 m is a tower-like structure composed of eight horizontal elements (named {\em floors}).
Each floor is a 8 m long marine grade aluminum structure connected to its next neighbours by means of eight tensioning ropes \cite{musumeci2006}.
The arrangement of these ropes is such that each floor is forced to a position
perpendicular to its vertical neighbours, as shown in Fig.~\ref{towersk}.
An iron anchor fixes the structure to the seabed while an appropriate buoyancy at the top provides the pull to keep the structure taut.
The floors are vertically spaced by 40 m, with the lowermost one positioned 100 m above the sea bottom.
Each floor holds  four Optical Modules (OM), two at each end, one looking
vertically downwards and the other horizontally outwards.
The structure is designed to be assembled and deployed in a compact configuration (Fig.~\ref{torre}) and to be unfurled once
on the sea bottom under the pull provided by the buoy.

\begin{figure} 
\begin{minipage}{\columnwidth}
\includegraphics[width=1.\textwidth]{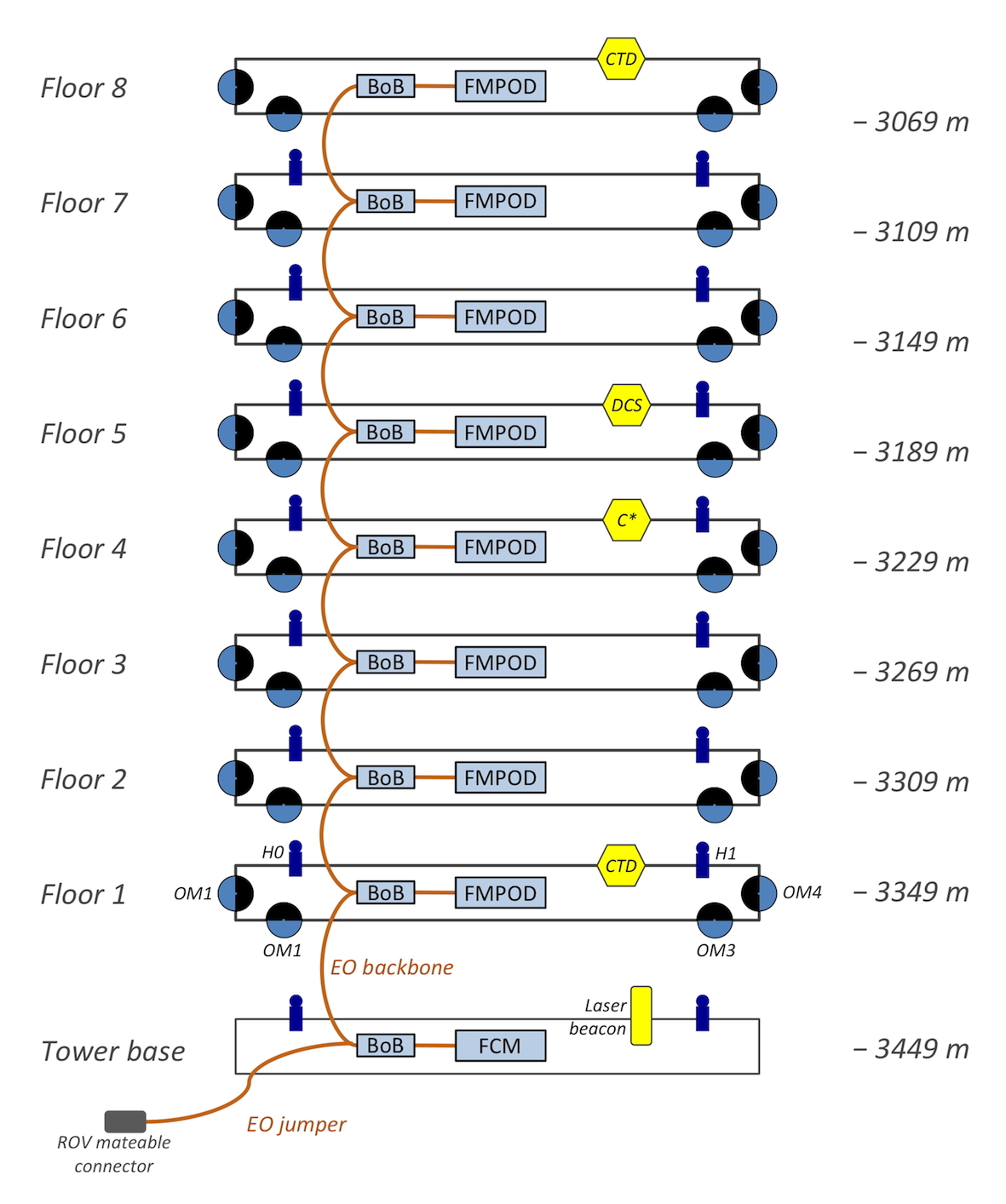}%
\end{minipage}
\caption{Schematics of the NEMO Phase-2 tower, including the backbone cabling (orange), the Floor Module Protective Oceanic Device (FMPOD) and the electro-optical breakouts (BoB). Connection to the Junction Box is provided through a ROV-mateable hybrid connector, installed in the tower base. The position of the environmental probes are also indicated as well as those of the hydrophones (H0 and H1) (see text for details). Nominal depths of each floor are given on the right.}
\label{towerscheme}
\end{figure}

A schematics of the tower is shown in Fig.~\ref{towerscheme}.
In addition to the 32 OMs the instrumentation installed includes several sensors for calibration and environmental monitoring.
In particular for positioning calibration purposes two hydrophones are installed close to the ends of each floor and two others on the tower base. 
Being the deep seawater a dynamical medium, monitoring of its oceanographic and optical properties 
during the detector operation is important since they can have an
impact on the detector performance. For that reason environmental 
probes are installed on the tower: two 
Conductivity--Temperature--Depth (CTD) probes \footnote{Sea Bird Electronics, 37-SM Micro-CAT.}, installed on the $1^{st}$ and $7^{th}$ floor;
a light transmissometre used for the measurement 
of blue light attenuation in seawater (C*); a Doppler Current Sensor
(DCS) used to monitor deep sea currents installed on the 5$^{th}$ floor.

Power distribution and data transmission along the tower is fulfilled by means of a ``backbone'' electro-optical cable. 
This is a lightweight umbilical subsea cable\footnote{Nexans S.A.}, carrying 10 electrical conductors and 12 optical fibres.

\subsection{The Optical Modules}

\begin{figure} 
\begin{minipage}{\columnwidth}
\includegraphics[width=0.5\textwidth]{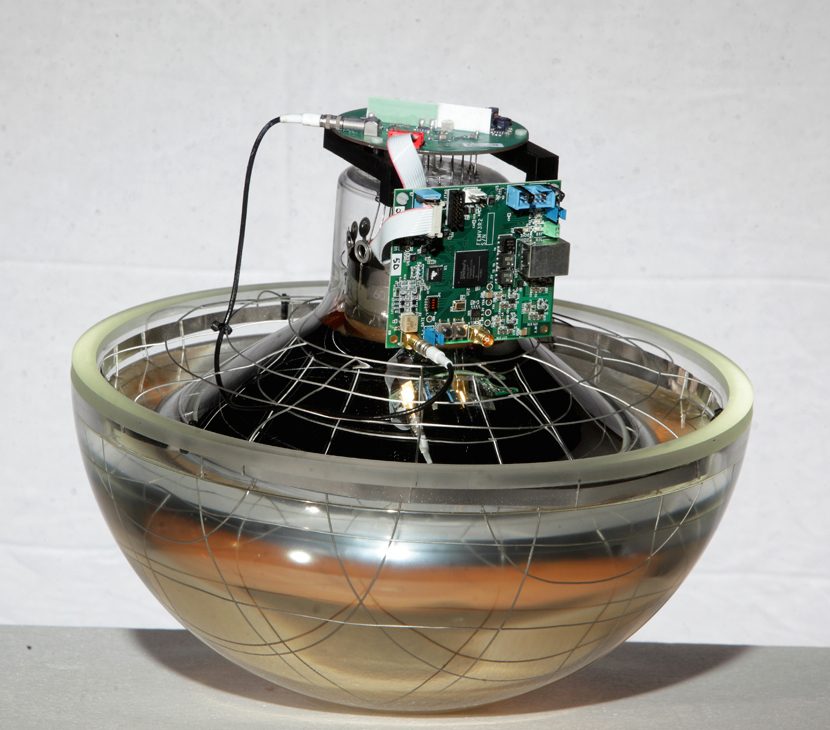}%
\includegraphics[width=0.5\textwidth]{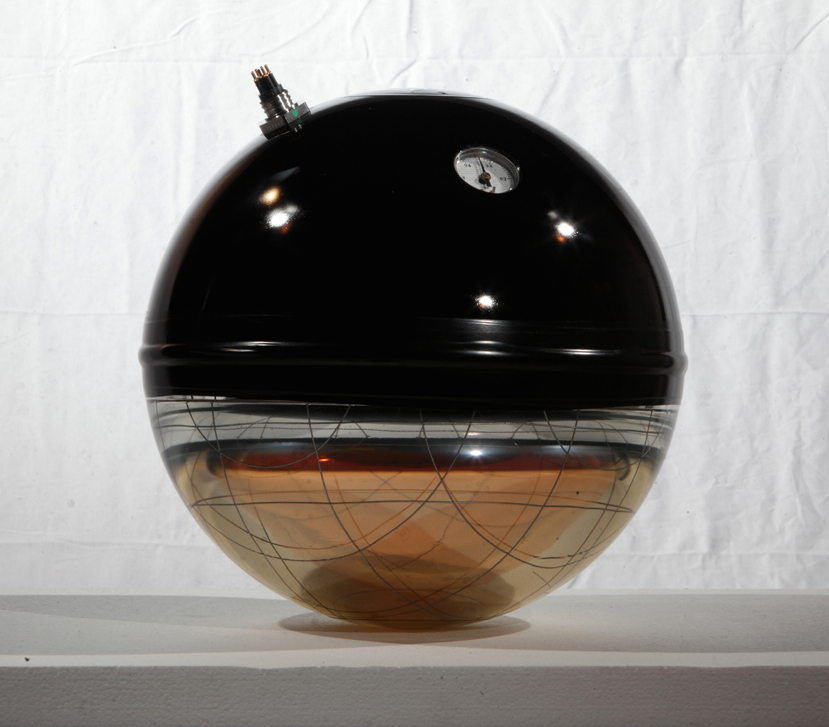}%
\end{minipage}
\caption{Left: One hemisphere of the optical module glass vessel containing the photomultiplier and the
front-end module board. Right: An optical module fully integrated with the electro-optical
feedthrough and the pressure-meter.}
\label{om}
\end{figure}

The OM is the basic element of the NEMO Phase-2 
detection unit \cite{aiello2013,leonora2013}. Each OM consists of a 13-inch pressure resistant (up
to 700 bar) borosilicate glass sphere\footnote{Nautilus Marine GmbH, VITROVEX flotation sphere.}, housing a
10-inch Hamamatsu photomultiplier (PMT R7081-SEL) \cite{aiello2010a} together with its high-voltage power supply, read-out electronics and calibration system (Fig.~\ref{om}).
Mechanical and optical contact between the PMT and the internal glass
surface is ensured by an optical silicone gel. A $\mu$-metal cage shields the
PMT from the Earth's magnetic field. 
A detailed description is given in \cite{aiello2013}.

PMT signal digitization is provided by a Front-End Module board that is also housed inside the OM \cite{nicolau2006}.
The board samples the analog PMT signal using two 8-bit Fast Analog to Digital Converters (Fast-ADCs) running at 100 MHz and staggered by 5 ns. This technique gives the
desired sampling rate yet allowing for a power dissipation lower than a single 
200 MHz ADC.
To match the large dynamic range of the PMT signal to the input voltage range of the ADCs, the signal is compressed by a non-linear circuit,
which applies a quasi-logarithmic signal compression.

Four OMs were also equipped with a prototype RFID based system that allowed to acquire oceanographic data with readout through the glass sphere without the use of feedthroughs \cite{cordelli2011}.

\subsection{Electronics and Cabling}

\begin{figure}
\begin{minipage}{\columnwidth}
\includegraphics[width=1.\textwidth]{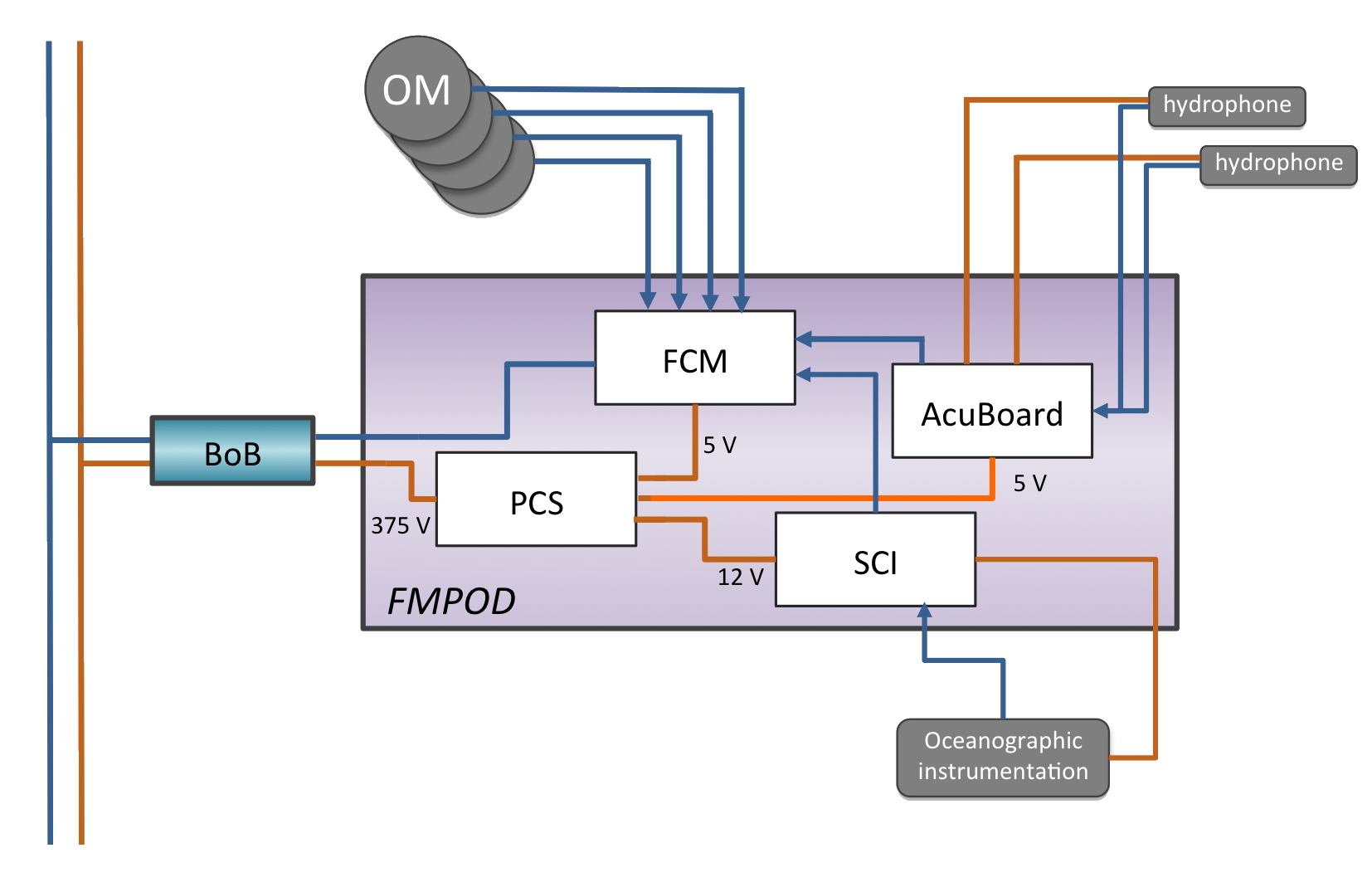}%
\end{minipage}
\caption{Schematics of the floor system. See text for the definition of the acronyms.}
\label{cablingscheme}
\end{figure}

A simplified scheme of the electronics and cabling of the tower is shown in Fig.~\ref{cablingscheme}.
At the level of each floor the backbone is split by means of breakout boxes (BoB). 
Each breakout is a high-density polyethylene vessel filled with
silicone oil and pressure compensated. The BoB is equipped
with two hybrid penetrators, which are used to split the backbone, and  two
connectors (one electrical and one optical) from the backbone to the floor cabling system. 

On each floor the four OM digital data are sent to a Floor Control Module board (FCM) housed in a  pressure resistant vessel, the Floor Module Protective Oceanic Device (FMPOD). The FMPOD is fitted in the middle of the floor mechanical structure and houses the data acquisition, control and the power distribution systems of the floor.
A similar setup is adopted for the tower base.

The 375 V supplied by the MVC is monitored at the level of the tower base and distributed to the eight floors. Inside the floor FMPOD a Power Contol System board (PCS)
provides conversion of the DC supply from 375 V to the low voltages needed by the electronics as well as monitoring and control of the main electrical parameters
of the floor.

A Slow Control Interface board (SCI) provides the interface to the oceanographic instruments installed on the floor via RS-232 serial standard.
In addition, each SCI has two analogue sensors to monitor humidity and temperature inside the FMPOD.

The acoustic board (AcuBoard) is a part of the positioning system described in section 3.5.1.

\subsection{Data Transmission System}

The link to the shore-station uses an optical fibre as
physical layer and implements a high speed serial link using a proprietary
data format \cite{ameli2008}. All data are encoded into a serial 800 Mb/s stream by a
serializer, converted into optical signal by an electro-optical transceiver
and transmitted to the shore station. In the communication protocol
the data stream is divided into 125 $\mu$s long frames of 10000 bytes each.

A transmission system through optical 
links based on  Dense Wavelength Division Multiplexing (DWDM)
technology was chosen \cite{damico2011}.
It is implemented by means of {\em add-and-drop}
passive devices that multiplex/demultiplex many optical channels at different
wavelengths into/from the same fibre. 
Two wavelengths are associated to each floor of the tower, one for the sea-to-shore and one for the shore-to-sea communication.
The optical
wavelengths are chosen according to the ITU standard grid with
100 GHz frequency spacing in the C-Band thus allowing up to 45 channels per fibre. 
The DWDM network provides, indeed, a {\em point-to-point}
communication between the shore-station equipment and the deep-sea
apparatus.
Each FCM contains an add-and-drop filter that allows to add or subtract 
the specific optical wavelength allocated to the floor.
Data from all floors are thus transmitted through the backbone in the
same fibre.

Detector data are received onshore by a dedicated electronics board, named Ethernet FCM (EFCM), which is based on
a Virtex-5\footnote{Xilinx Inc.} development board. This board collects the data received by
the underwater electronics and transfers them to the DAQ and storage systems through a Gigabit
Ethernet connection.

\subsection{Data Acquisition System}

Thanks to the extremely large communication bandwidth available (2 Gpbs), no hardware triggers 
are implemented underwater: all the digitized signals are sent to shore.
The data stream originated by each FCM is addressed on shore to the twin EFCM board, which transfers the data to the onshore Trigger and Data Acquisition System (TriDAS) \cite{chiarusi2011}.

Each detected photon pulse is sampled by the FEM and arranged by the FCM in a hit 
record with a mean size of 28 bytes. Consequently the tower averaged optical throughput is about 250 Mbps. 
The total amount of data from offshore includes also a stream of about 80 Mbps from the acoustic sensors and a negligible contribution of slow control data.
The electronics were designed to deal with an optical signal up to 150 kHz continuous
single rate on each PMT without dead time.

From the EFCMs, the PMT optical data-stream is routed through a 1 Gb Ethernet network to the first layer of the TriDAS, composed of two Hit Managers (HM) processes running on two CPUs. 
Each HM gathers data from half a tower and coherently time-slices the
continuous stream of data into time intervals of 200 ms. 
All the data corresponding to a given interval of time are sent by the HMs to a single TriggerCPU process (TCPU). 
Subsequent time slices are addressed to the others TCPU processes,
implementing the trigger algorithms for background rejection. 
A reduced selected stream is then addressed, through a 1 Gb Ethernet switch, from all the TCPUs to the Event Manager (EM) server which is deputed to write the trigger selected data on the local temporary storage device. 
Finally data are copied to the long lasting storage facility at the Laboratori Nazionali del Sud in Catania (Italy) by means of a dedicated 1 Gb Ethernet point to point connection.

\subsection{Positioning and calibration}

\subsubsection{Acoustic positioning system}

Since the tower structure is not rigid, floors are subject to the effects of deep-sea currents than can distort the vertical line shape of the tower, making the floors rotate and tilt.
For a proper reconstruction of the muon tracks, which is based on space-time correlation of Cherenkov photons hitting the OMs, the knowledge of the position of each OM with a precision of the order of  10 cm is needed. This can be obtained by using a system based on acoustic triangulation as demonstrated by previous experience gained with the NEMO Phase-1 prototype \cite{aiello2010}.

The acoustic positioning system installed on the NEMO Phase-2 tower consists of couples of hydrophones mounted in fixed positions close to the end of each floor (H0 and H1 in Fig.~\ref{towerscheme}) \cite{viola2013,viola2014}.
For testing and validation purposes different types of hydrophones were used.
The six lowermost floors are equipped with large broadband hydrophones (10 Hz - 70 kHz)\footnote{Model SMID TR-401.}.
A couple of free flooded rings (FFR) hydrophones\footnote{Sensor Technology Ltd.} is installed on the $7^{th}$ floor.
Finally, two of the OMs on the $8^{th}$ floor are equipped with custom designed piezoelectric acoustic sensors glued to the internal surface of the glass sphere.
Data from all acoustic receivers
are sampled offshore at a frequency of 192 kHz with a 24 bit resolution and 
continuously sent to shore.
Offshore data are time stamped with the absolute acquisition time by a dedicated synchronous system
which is phased with a GPS time station\footnote{Symmetricom XLi.}.
The acoustic position system is completed with a set of autonomous acoustic beacons installed on the seafloor at a distance of approximately 400 m from the tower and one beacon at the base of the tower.

An independent real time monitoring of the floor orientation is provided 
by an Attitude Heading Reference System (AHRS), that includes triaxial gyroscopes, accelerometers and magnetometers,
placed inside each FMPOD.

\subsubsection{Time calibration System}

The reconstruction of physical events (e.g., muon tracks) requires sub-nanosecond synchronisation between the OMs. 

The synchronous communication protocol implemented on the tower provides synchronization with a unique clock source (Master Clock) for all electronics boards and OMs. The clock distribution system is based on a GPS station, which provides the absolute time, encoded in IRIG-B 100-1344 standard, and a high-stability 10 MHz clock which works as Master Clock for the detector \cite{circella2009}. 

Due to the clock propagation delay from shore to the tower, each OM has its own time offset  to be measured and compensated.
This is done using a fast light pulser installed in each OM.
The light pulsers are controlled by dedicated FPGA\footnote{Altera Cyclone III.} based control boards located in each FMPOD, which also allow setting the intensity and the frequency of the flashes of each light pulser.
The system is operated from shore.
The control boards can also perform measurements of the propagation delays of the commands to reach each of the light pulsers.
The propagation delays have been proven to remain stable down to the level of the resolution of the measurements, i.e. within 100 ps.

Prior to the deployment, a calibration of the OMs of the full tower has been performed by using an external system exploiting a fast laser source (Hamamatsu PLP-10) coupled to a network of optical fibres of calibrated lengths. Multi-mode, graded index optical fibres have been used for this application.

In order to have redundancy in time calibration, for exploring new solutions proposed for a km$^3$-size detector as well as for measurements of the optical properties of water, the NEMO Phase-2 tower is also equipped with additional calibration devices comprising small LED pulsers, installed inside selected OMs in such a way to illuminate the OMs of the upper floors, and a laser beacon installed on the base of the tower.

\section{Installation and operation}

The NEMO Phase-2 tower was deployed on March 23 2013 during a sea campaign 
operated by the Multi Service Vessel `Nautical Tide'\footnote {MTS/FUGRO Chance Company}. 
The tower deployment operations took approximately 6 hours.
The tower was first lowered to the seabed and then connected to the
CTF by means of a surface controlled ROV.
After the connection the tower was unfurled reaching its final configuration.
Data acquisition started immediately.

During the same operation one autonomous acoustic beacon was also deployed at about 400 m North of  the tower.
A second one was installed in a separate sea campaign on July 20 2013 at about 400 m East of the tower.
Together with the one installed on the tower base they form the acoustic Long Baseline (LBL) system.  
These beacons were set to a repetition rate of 0.5 Hz and were not synchronized to the GPS 
clock distributed from shore. Their time of emission had to be 
re-synchronised with respect to the detector clock by means of the monitoring hydrophone mounted
on the tower base in a fixed and known position.
A malfunctioning in the power line cable of the monitoring hydrophones prevented the possibility of calibrating the LBL system.
Consequently it was not possible to determine the position of each single hydrophone of the tower.
However, information on the vertical position of the tower was provided by the CTD probes
which were found to be at the expected mean depths of 3358 m ($1^{\rm st}$ floor) and 3118 m ($7^{\rm th}$ floor).
This information was checked by measuring the time delays of acoustic signals, emitted by the beacon installed on the tower base, between the $1^{\rm st}$ floor and the other floors and found to be consistent with the CTD data.

\begin{figure} 
\begin{minipage}{\columnwidth}
\includegraphics[width=1\textwidth]{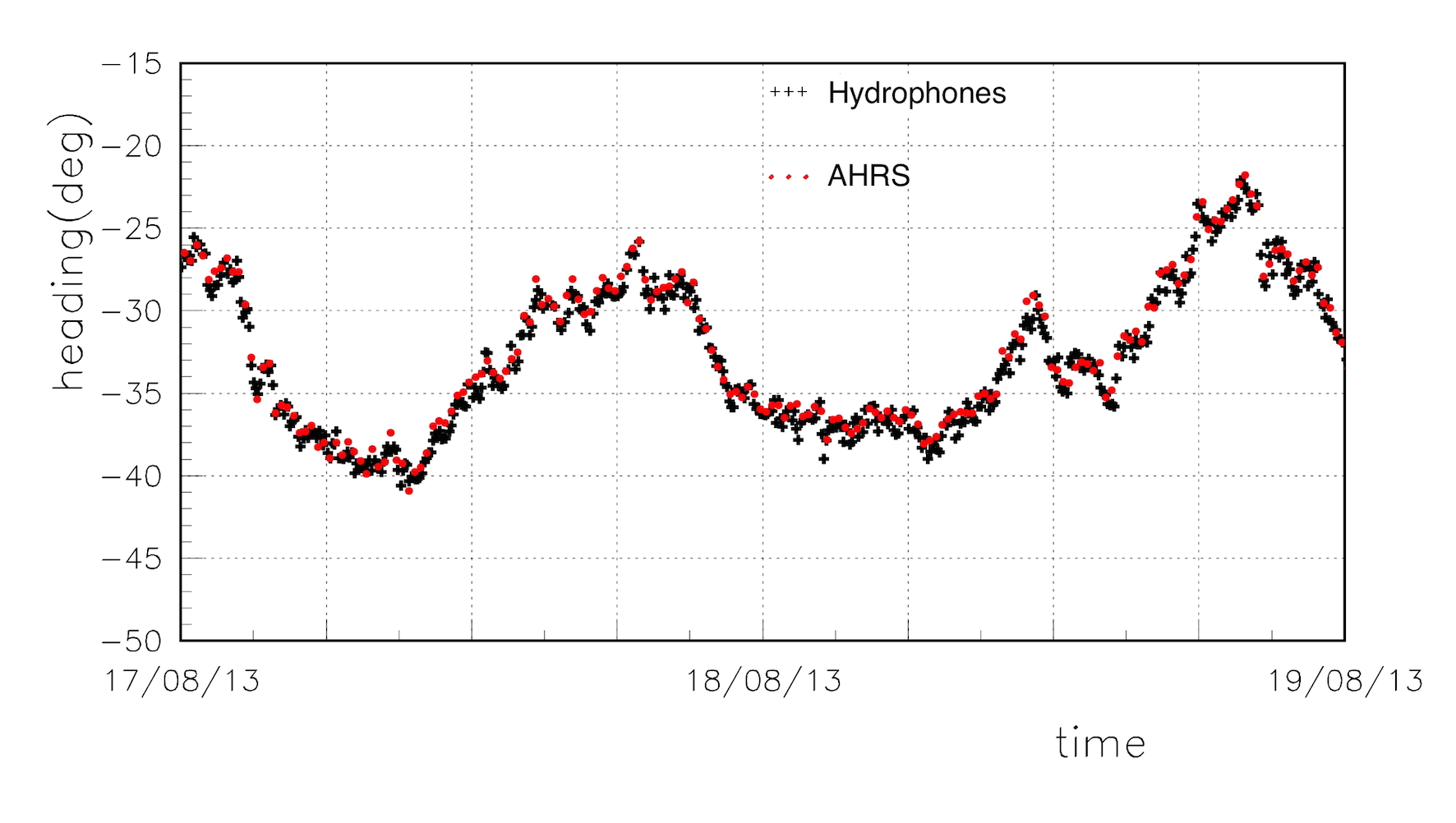}%
\end{minipage}
\caption{Orientation of Floor 6 determined using the hydrophone data (black crosses) compared to the heading information determined by the AHRS installed on the Floor Control Module.}
\label{bussole}
\end{figure}

Although in a reduced configuration, the acoustic data allow to recover 
the orientation (heading) and pitch and roll of each floor with a 
precision better than 1 degree.
Using a global fit algorithm, the position of each hydrophone was calculated with an estimated error of about 1 m.
The acoustic data were combined with depth information from the two CTDs.
A fixed distance between floors of 40 m and a 5 m uncertainty on the position of the 
LBL beacons, determined by the ship GPS system at the time of deployment, were assumed.
As an example, a floor heading reconstruction by using the hydrophones system
is shown in Fig.~\ref{bussole} for period of two days. 
This is in excellent agreement with the AHRS
heading reconstruction (15 minutes averages) normalized to the hydrophone data.

Despite the lack of a precise positioning of each single OM it was however possible to reconstruct downgoing muon tracks and determine
the Depth Intensity Relation of muons in water as reported in \cite{aiello2015}.

In a limited number of runs a new test device, i.e. a beacon located at the 
tower base whose emission is synchronous with the master clock, 
was used. In this case it was possible to reduce the 
effect of systematic errors and determine  the 
position of each hydrophone with an accuracy of about 30 cm comparable to those obtainable with a complete LBL.

The commissioning phase of the tower lasted up to 30 June 2013. During this phase the PMT gains were adjusted several times to obtain the same gain factor of $5 \times 10^7$.
This converts each single photon electron (s.p.e.) into a mean integrated charge
of about 8 pC.
Thresholds were adjusted at approximately 0.25 s.p.e.
As a consequence data rates reported in the following have to be considered stable only after the completion of this phase.

Trigger used during the data taking are described in \cite{aiello2015}. 

\section{Long term monitoring of the optical background}

During seventeen months of operation the NEMO Phase-2 tower allowed for a continuos long term monitoring of the site characteristics.
In particular, the optical background rate, due to the presence of Cherenkov light produced in the decay of  $^{40}$K as well as light emitted by
bioluminescent organisms, was studied.

\begin{figure} 
\begin{minipage}{\columnwidth}
\includegraphics[width=1\textwidth]{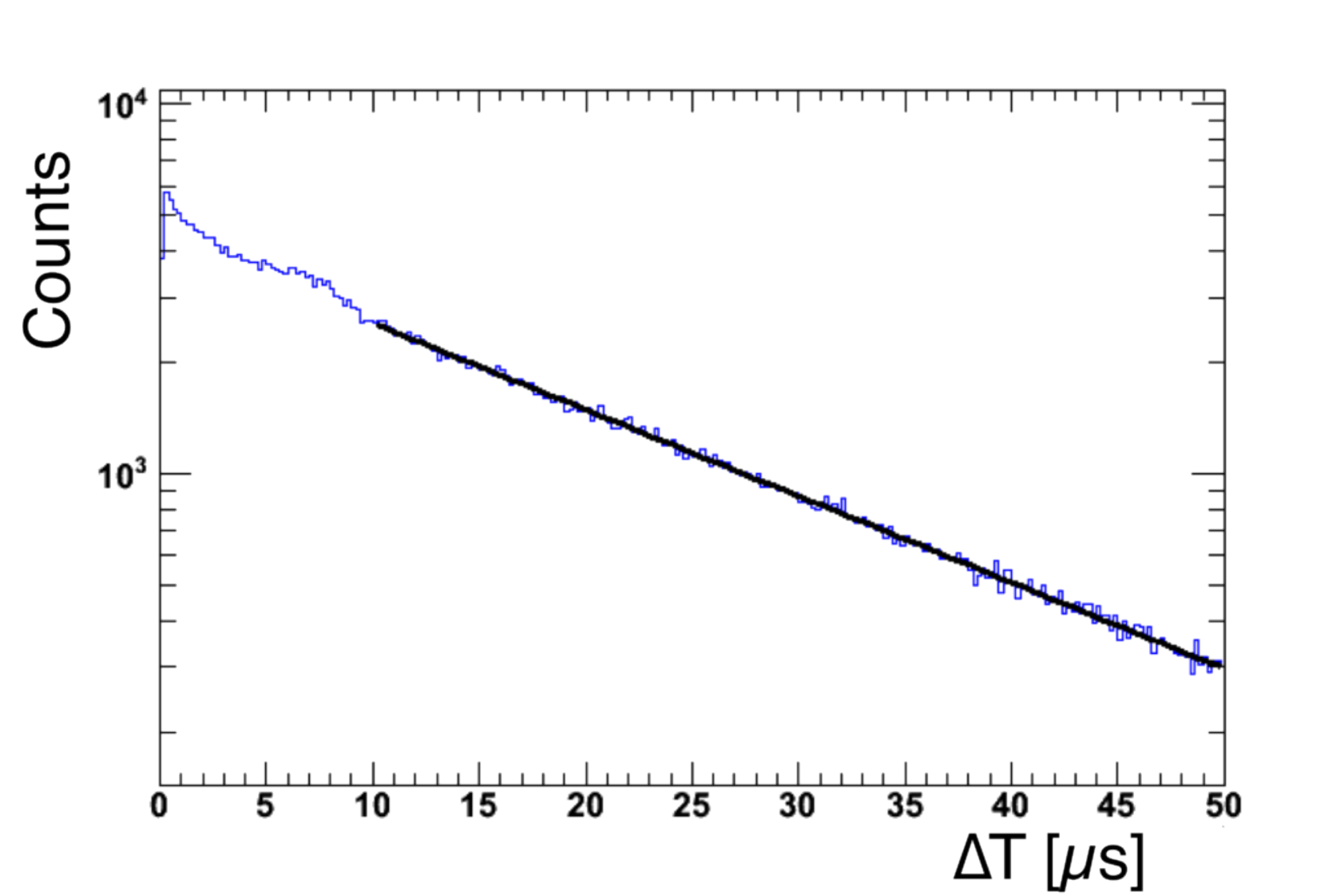}%
\end{minipage}
\caption{Time difference between two consecutive hits.
The exponential fit above $\Delta t = 10 \mu$s (black line) gives a characteristic time $\tau = 1.8 \times 10^{-5}$ s.
The bump at $\Delta t \approx 7 \mu$s is due to the presence of PMT afterpulses. For $\Delta t < 1 \mu$s the effect of electronics deadtime is present.}
\label{deltaT_distribution}
\end{figure}

The distribution of the time difference between successive hits, $\Delta t$, of a typical OM, is shown in Fig.~\ref{deltaT_distribution}.
It has the expected exponential form, with a slope of $\tau = 1.8 \times 10^{-5}$ s corresponding to a purely random background singles rate of about 55 kHz. 

The expected singles rate has been evaluated by using a complete GEANT4 simulation, that accounts for the $^{40}$K contribution, the PMT dark counts and the presence of radioactive materials in the glass sphere \cite{hugon2015}. The two latter were measured in the laboratory. The water absorption length of the KM3NeT-It site has been used \cite{riccobene2007}.
The simulation also takes into account the shadowing effects of the mechanical structure holding the OMs.
The result of this simulation gives a value of $54 \pm 3$ kHz in agreement with the in situ observed rate.
Optical module relative efficiencies were measured in the laboratory on a set of six optical modules and used as input to tune the Monte Carlo simulations.
The major source of uncertainty on the simulated singles rate comes  from the spread of the quantum efficiencies if the photocatodes examined  \cite{hugon2015} and can be considered as representative of the full set of PMTs used in the NEMO Phase-2 tower. The standard deviation of these measurements was taken into account and propagated to the simulation results.

Double coincidences, originating from $^{40}$K decays, between close neighbour OMs have also been studied. As an example for the two OM couples at the extremities of the 1$^{\rm st}$ floor, values of $20.9\pm0.9$ Hz and $21.5\pm0.8$ Hz were  measured. For the other OM couples the values range from 19.0 Hz to 22.5 Hz with an average value of $20.45\pm1.5$ Hz. The spread can be mainly ascribed to differences in the PMT threshold and on the photocatode quantum efficiency.
For OM couples with the same geometry the Monte Carlo simulations gives a value of 21.6 Hz  \cite{hugon2015} in good agreement with the measured data.

\begin{figure} 
\begin{minipage}{\columnwidth}
\includegraphics[width=1\textwidth]{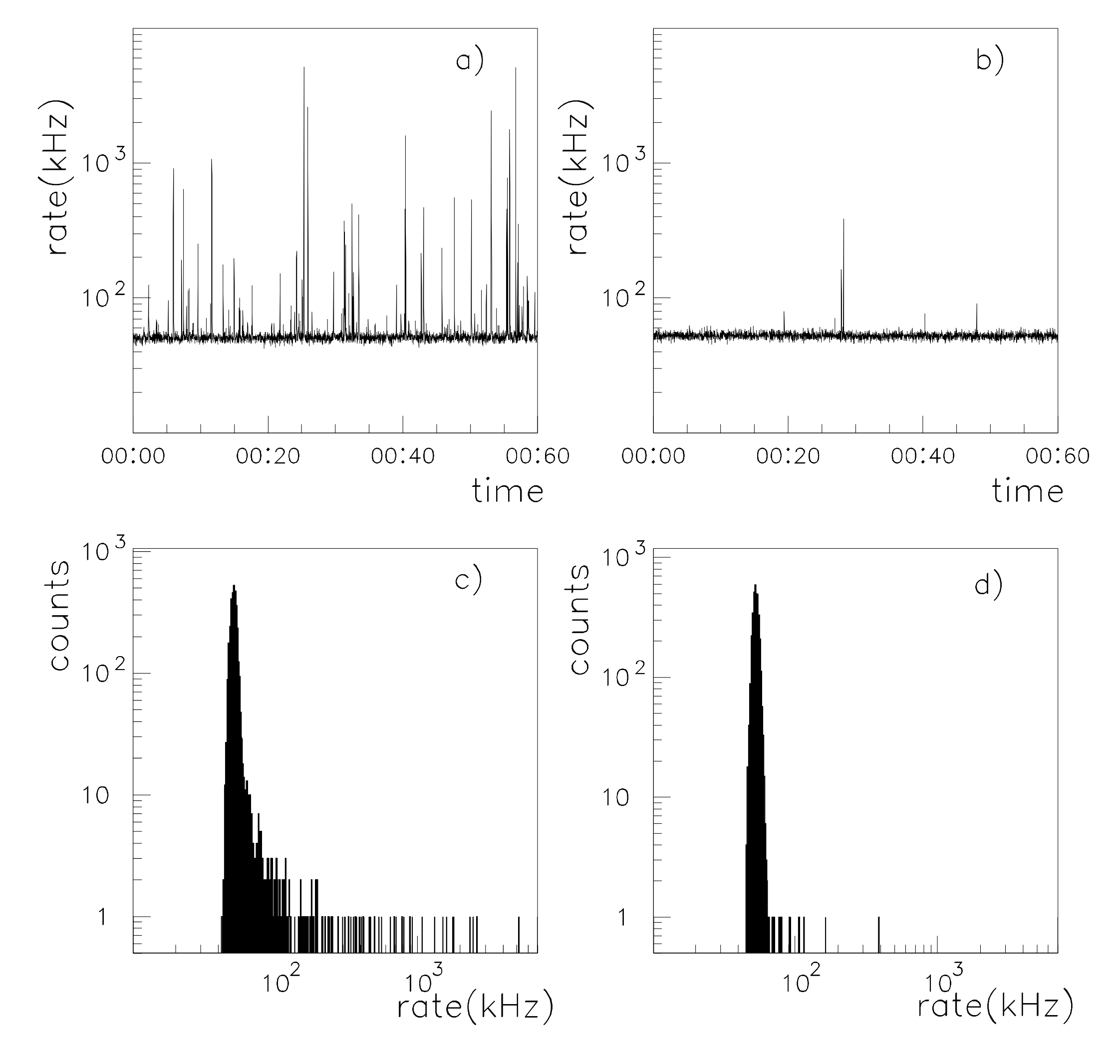}%
\end{minipage}
\caption{Singles rate as measured in the monitoring channel as a function of time for one hour time slots measured on July 11 2013 (panel a) and on September 11 2013 (panel b). Panels c) and d) show the corresponding histograms of the monitored rate values.}
\label{figrate1h}
\end{figure}

The rates determined as in Fig.~\ref{deltaT_distribution} on the acquired data suffer from the dead time introduced by the data acquisition system that limits the maximum data rate transferable to about 150 kHz.
However, the FEM board on each OM provides an independent monitoring of the average count
rate by counting once per second the number of hits with amplitude exceeding a
threshold of  $\approx0.25$ s.p.e. in a 10 ms time window. 
This estimate does not suffer from dead time limitations allowing to monitor the
signal rate up to about 6.5 MHz.
Figure \ref{figrate1h} shows the distribution of the singles rates (measured by the OM rate monitor) in one OM for two one-hour time slots taken in different periods with different optical background characteristics. In the top panels (a and b) the rate is shown as a function of time. In the bottom panels (c and d) the corresponding histograms of the rate distributions are shown.
In general, the instantaneous rate exhibits a flat baseline at around 55 kHz with some sporadic bursts.
The baseline corresponds to
the contribution of the $^{40}$K discussed above while light bursts can be attributed to the presence of bioluminescent organisms.
The amount of these bursts can vary in time, as shown by the two examples in Fig.~\ref{figrate1h}.
To monitor these variations, as well as possible variations in the baseline rate, 
histograms like those of Figs.~\ref{figrate1h}c)-d) have been built for 15 minute-long time frames covering the whole operation period.
From each histogram the baseline rate was extracted as median of the distribution.
The amount of light bursts was determined as the fraction of time with rate exceeding an arbitrarily fixed threshold of 100 kHz (burst fraction).

\begin{figure} 
\begin{minipage}{\columnwidth}
\includegraphics[width=1\textwidth]{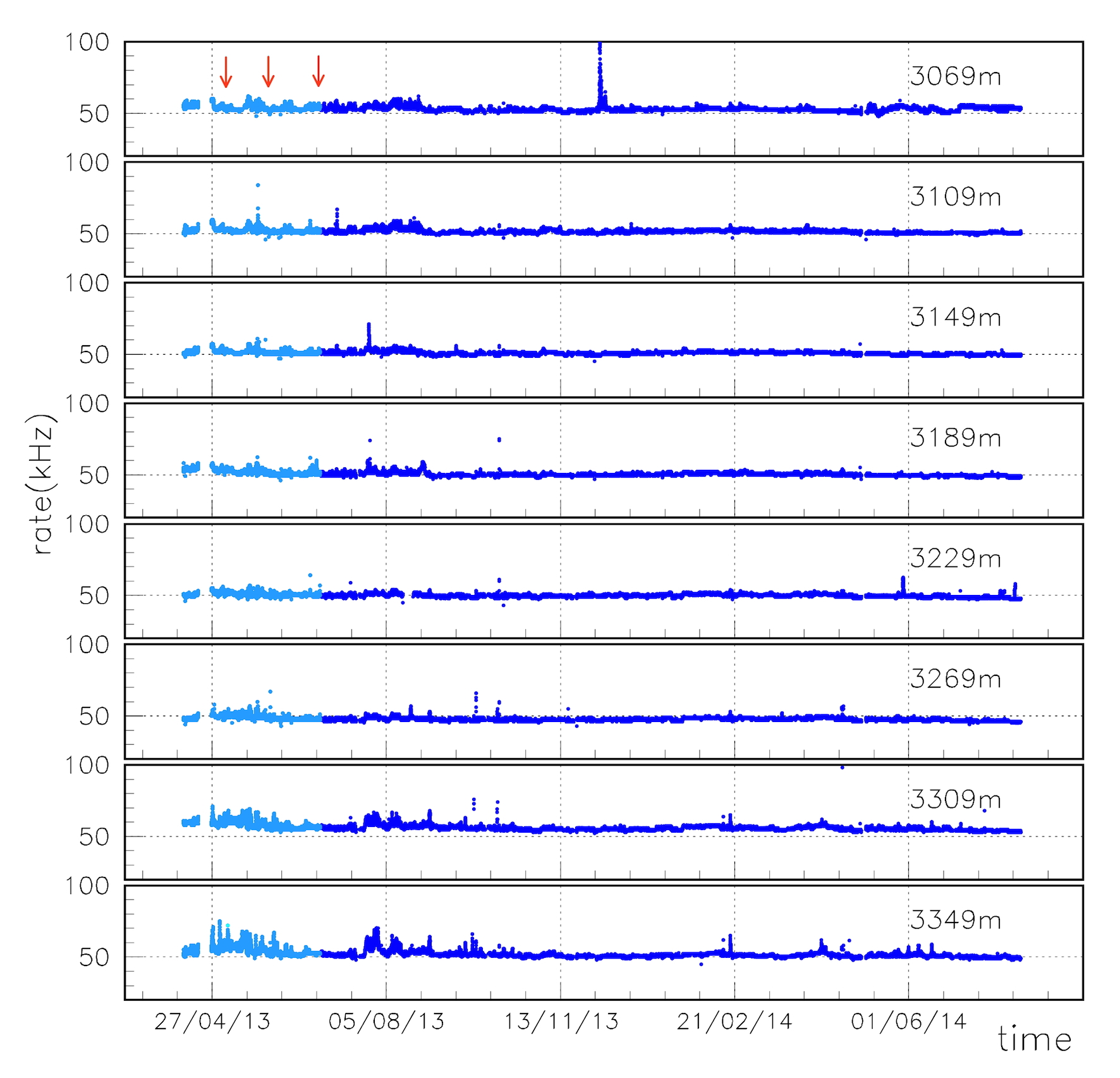}%
\end{minipage}
\caption{Baseline rate measured for the whole operation period of the tower at different depths. All selected PMTs are down-looking. Light blue dots correspond to the commissioning period of the detector.  The arrows indicate the dates when PMT high voltage adjustments were performed.}
\label{figratef1}
\end{figure}

In Fig.~\ref{figratef1}, the baseline rate for eight OMs, all chosen with the same down-looking orientation and located at different depths, is shown for the whole operation period of the tower.
For the sake of completeness the data taken during the commissioning period are also shown (light blue dots).
The observed baseline rates do not show large fluctuation above 50-60 kHz and are in agreement with previous measurements performed in the same site.
Sporadic high-baseline rates, which are however always below 100 kHz, are observed. They can be attributed to periods with a higher level of bioluminescent activity.
The average value for the whole measurement period and for all floors is compatible with the estimate of the $^{40}$K decay contribution. No significant effect due to the depth is observed.
The measured rates are in general lower and more stable than those observed in the ANTARES neutrino detector site \cite{ageron2011} and reported in \cite{tamburini2013}.

\begin{figure} 
\begin{minipage}{\columnwidth}
\includegraphics[width=1\textwidth]{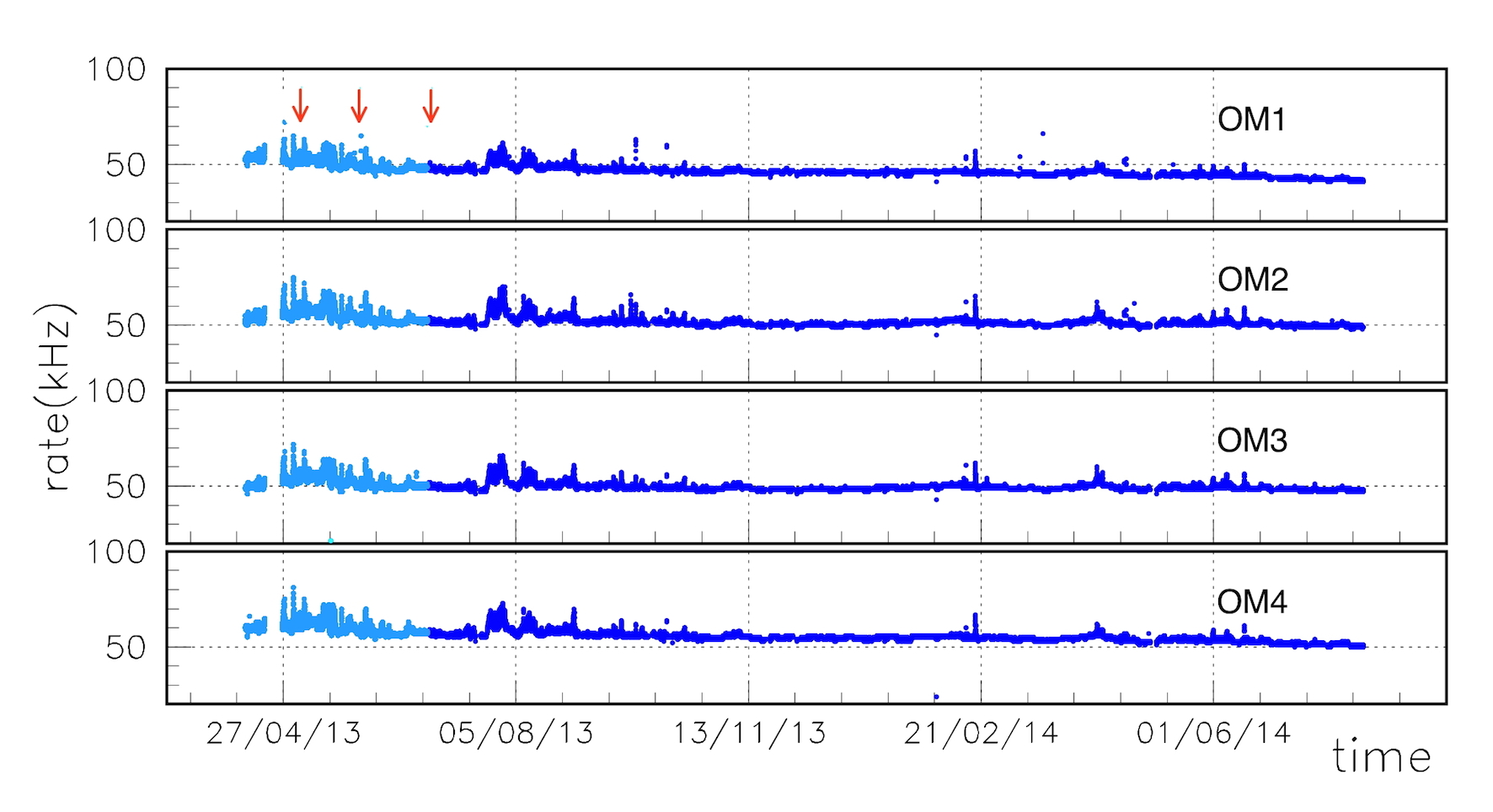}%
\end{minipage}
\caption{Baseline rate measured for the whole operation period of the tower for the four OMs of the 1$^{\rm st}$ floor of the tower.
The OM numbering follows the scheme illustrated in Fig.~\ref{towerscheme} with OM 1 and 4 that are horizontally looking and OM 2 and 3 downward looking.
Light blue dots correspond to the commissioning period of the detector.  The arrows indicate the dates when PMT high voltage adjustments were performed.}
\label{figratef2}
\end{figure}

For a comparison between optical modules at the same depth but with different orientation, the baseline rates for the for OMs of the 1$^{\rm st}$ floor of the tower are shown in Fig.~\ref{figratef2}.
Also in this case the rates are similar. However, a slight decrease with time of the rates measured with the two horizontally looking OMs can be observed.
The origin of this is unknown but can be reasonably attributed to the deposition of a thin sediment layer on the uppermost part of the optical module.

\begin{figure}
\begin{minipage}{\columnwidth}
\includegraphics[width=1\textwidth]{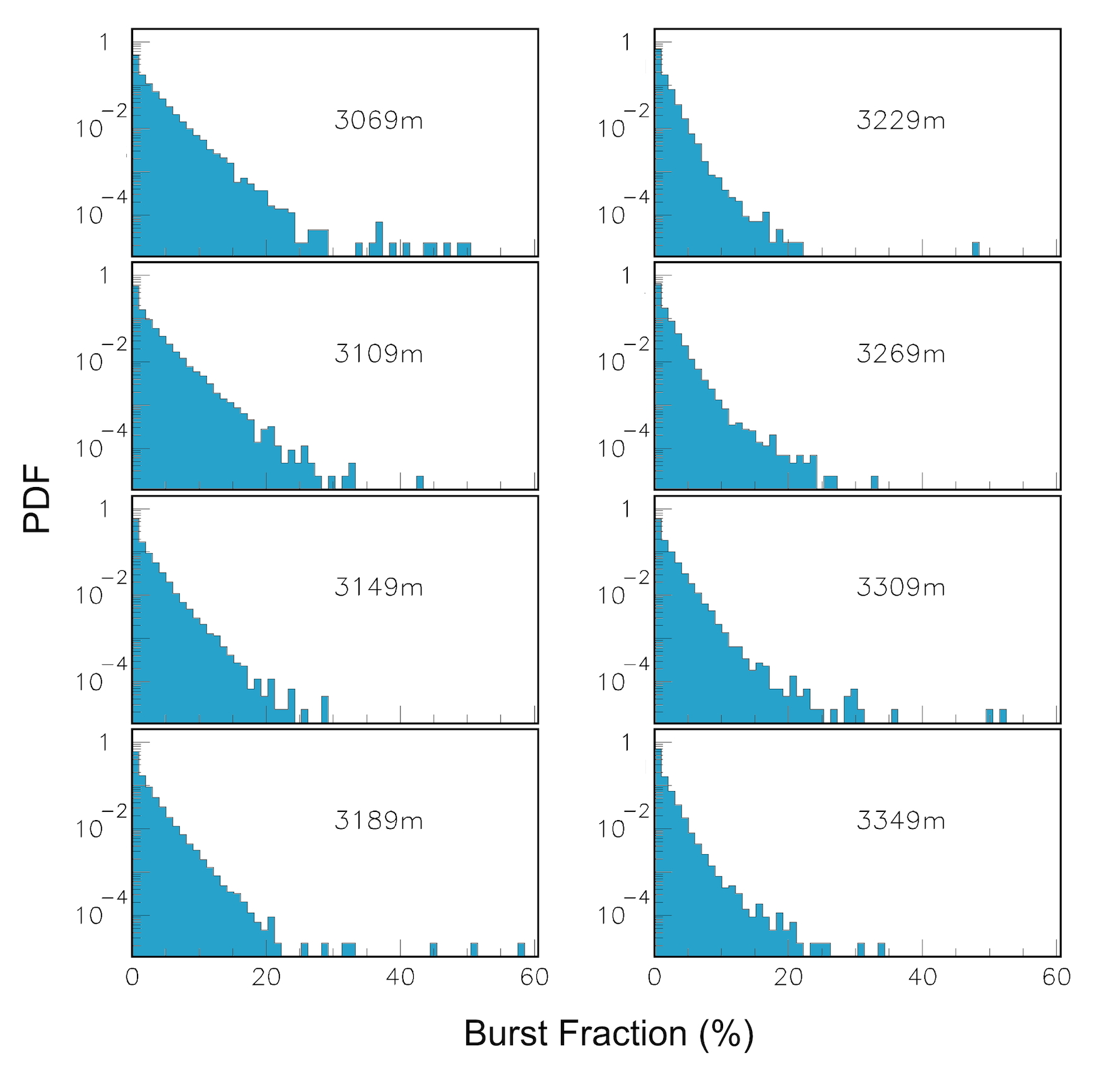}%
\end{minipage}
\caption{Probability Density Function (PDF) of observing a given burst fraction, defined as having a rate exceeding 100 kHz in a 10 ms time window, in eight down-looking OMs placed at different depths.}
\label{figbff}
\end{figure}

\begin{figure} 
\begin{minipage}{\columnwidth}
\includegraphics[width=1\textwidth]{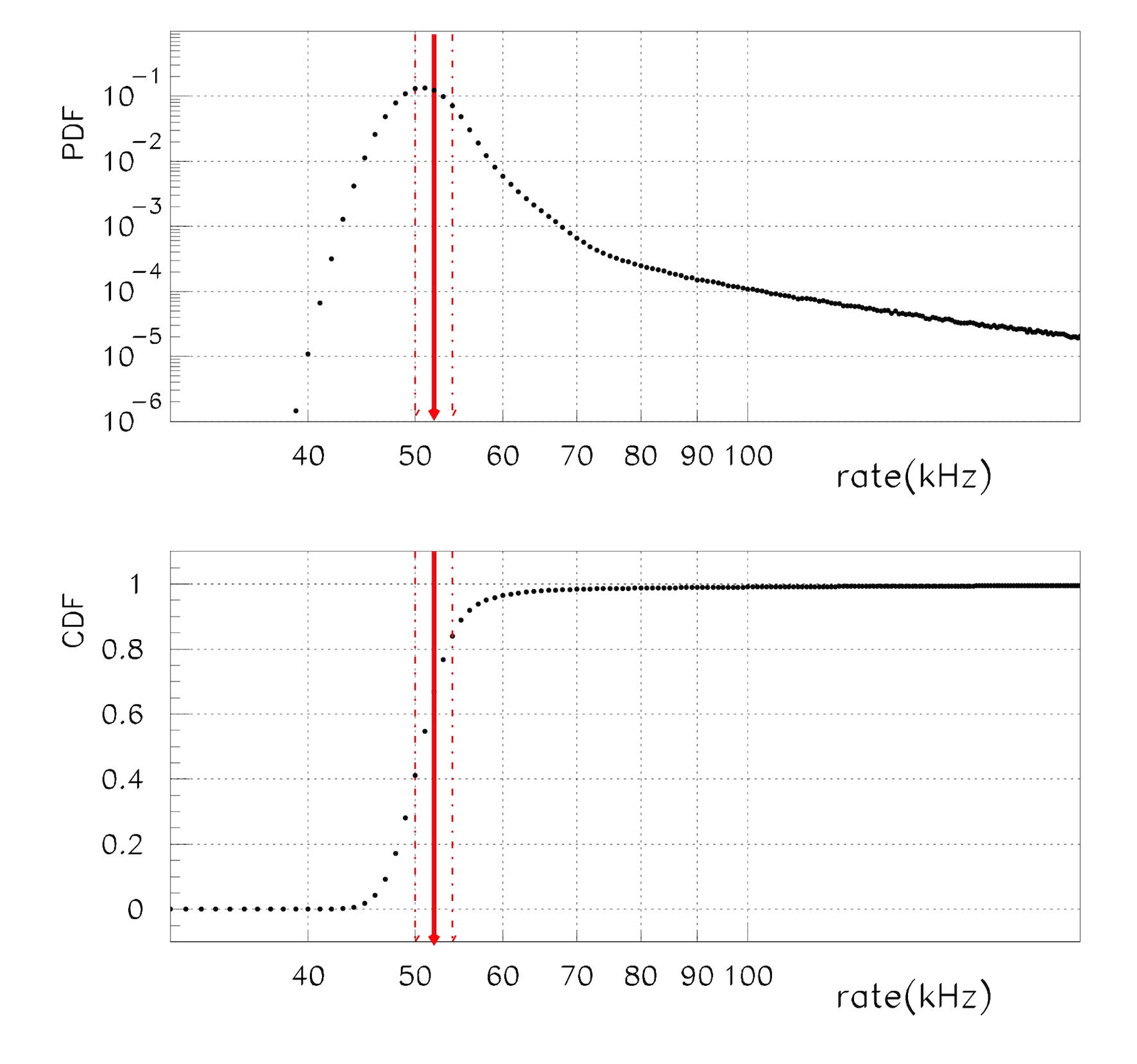}%
\end{minipage}
\caption{Probability Density Function (PDF) and Cumulative Density Function (CDF) of singles rate (black dots) for the whole set of data taken from July 2013 to August 2014 of one PMT. The solid red line indicates the result of the GEANT4 simulation described in the text. Dashed red lines give the error in this simulation.}
\label{pdf}
\end{figure}

For the same set of eight down-looking OMs the Probability Density Function of having a rate exceeding 100 kHz in a 10 ms time interval are reported in Fig.~\ref{figbff}.
No significant depth dependence is observed.
This effect can be seen also in Fig.~\ref{pdf} (upper panel) where the Probability
Density Function of the rate has been calculated after the last HV adjustment for the period going from July 2013 until August 2014. 
In particular, the contribution of bioluminescence bursts is shown for an optical module located on the $1^{\rm st}$ floor of the tower.
For the examined OM, the rate distribution is peaked around 50 kHz and becomes negligible above 100 kHz. The other optical modules show very similar behaviors with peaks located in the range 45-55 kHz. The Cumulative Density Function of the rate has also been calculated and it reaches the value of 0.99 in the range 100-200 kHz. 
This leads to the important conclusion that the probability to have the TriDAS in dead time is less than 1\% all along the full year observation. There is no period in which the OMs needed to be switched off for high bioluminescence activity.

\begin{figure} 
\begin{minipage}{\columnwidth}
\includegraphics[width=1\textwidth]{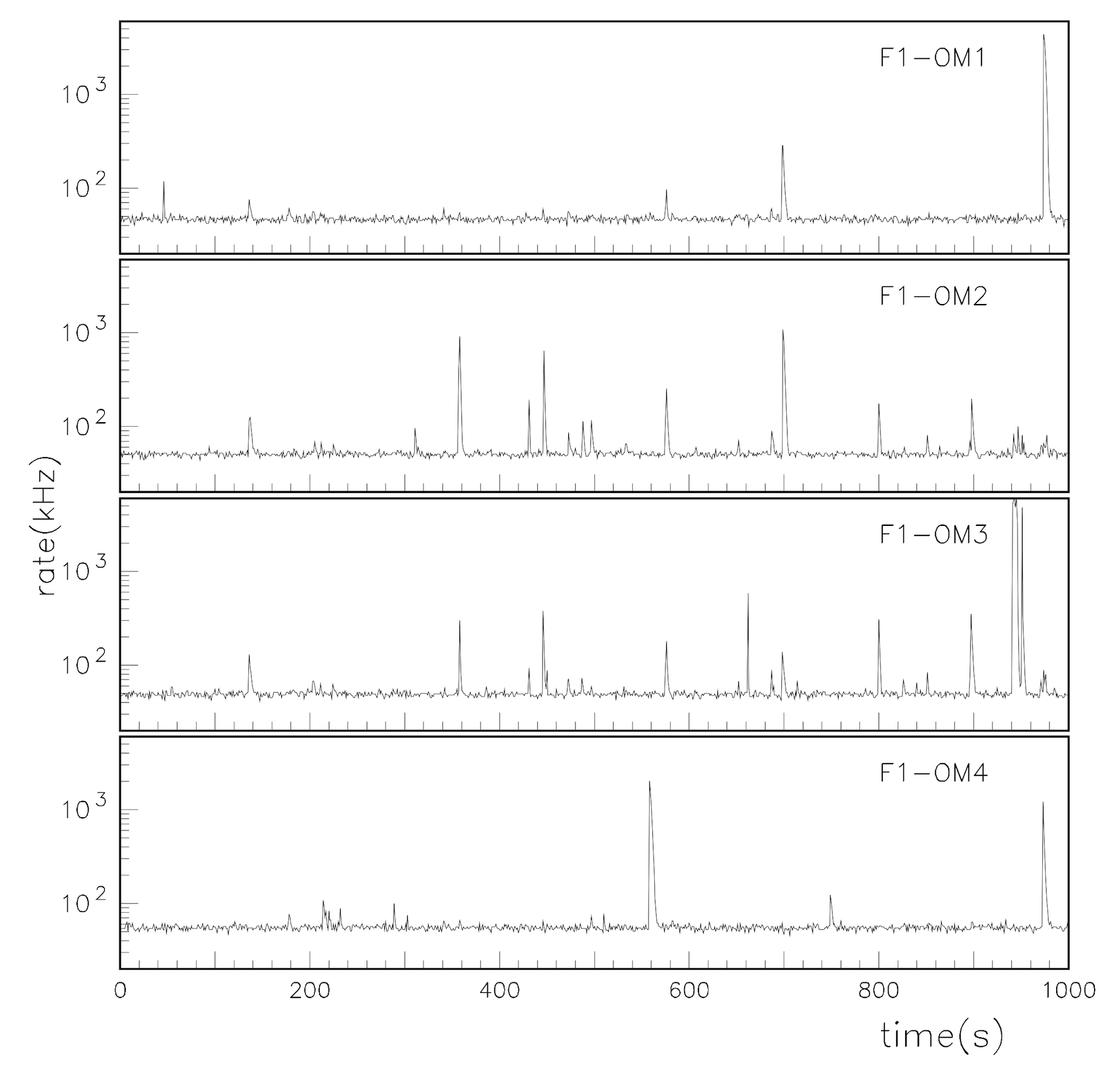}%
\end{minipage}
\caption{Singles rate as measured in the monitoring channel of the four OMs of the 1$^{\rm st}$ floor as a function of time for a 1000 s time slot measured on July 11 2013, corresponding to a period of relatively high-level bioluminescence activity (same as the one shown in Fig.~\ref{figrate1h}a).
}
\label{ratesnapshot}
\end{figure}

For a better evaluation of the impact of bioluminescent bursts on the detector performance a study of the coincidences between bursts on OMs on the same floor has been carried out.
For a visual comprehension a sample snapshot of 1000 s, recorded during a period of high level bioluminescence, is shown in Fig.~\ref{ratesnapshot}. The four panels show the time series of the instant rates measured on the four OMs of the 1$^{\rm st}$ floor.  The OM numbering follows the scheme illustrated in Fig.~\ref{towerscheme}.
A quantitative analysis has been performed by evaluating over the whole data sample the probability $P_i$ of having a rate exceeding 100 kHz in a single OM and the probability $P_{ij}$ of having a rate exceeding 100 kHz simultaneously in a pair of OMs. This has been computed for all the six possible pairs of OMs of the 1$^{\rm st}$ floor
and reported in Table~\ref{probab}.
The correlation factor, defined as
$r_{ij} = (P_{ij} - P_i \times P_j) / \sqrt{(P_i - P_i^2)(P_j - P_j^2)}$
is also reported.
Close-by OMs (pair 1,2 and 3,4) and the two downward looking OMs (pair 2,3), which observe the same volume of water, show a higher rate of burst coincidences with a higher degree of correlation. However, the probability of observing the same burst simultaneously in both OMs  is always less than 0.2\%.
For distant OMs the probability of simultaneously observing the same burst is of the order of $10^{-3}$ or less with a lower degree of correlation.

\begin{table}
\caption{\label{probab}Probabilities $P_i$ and $P_j$ to observe independently a rate exceeding 100 kHz in a pair $(i,j)$ of OMs compared with the probability $P_{ij}$ to observe a burst exceeding 100 kHz simultaneously in both OMs. A correlation factor $r_{ij}$ (see text) is also reported.
The data refers to the six possible pairs of OMs of the 1$^{\rm st}$ floor.}
\begin{tabular}{ccccc}
OM pair & $P_i$ & $P_j$ & $P_{ij}$ & $r_{ij}$\\ \hline
(1,2) & $5.03 \times 10^{-3}$ & $1.09 \times 10^{-2}$ & $2.55\times 10^{-3}$&$0.34$\\
(3,4) & $9.30 \times 10^{-3}$ & $5.71 \times 10^{-3}$& $2.67\times 10^{-3}$& $0.36$ \\
(1,3) & $5.03 \times 10^{-3}$ & $9.30\times 10^{-3}$& $1.38 \times 10^{-3}$& $0.20$\\
(2,4) & $1.09 \times 10^{-2}$ & $5.71\times 10^{-3}$& $1.46 \times 10^{-3}$& $0.18$\\
(2,3) & $1.09 \times 10^{-2}$ & $9.30\times 10^{-3}$ & $4.33\times 10^{-3}$& $0.42$ \\
(1,4) & $5.03 \times 10^{-3}$ & $5.71\times 10^{-3}$ & $2.70\times 10^{-4}$& $0.05$\\
\end{tabular}
\end{table}

\begin{figure}
\begin{minipage}{\columnwidth}
\includegraphics[width=1\textwidth]{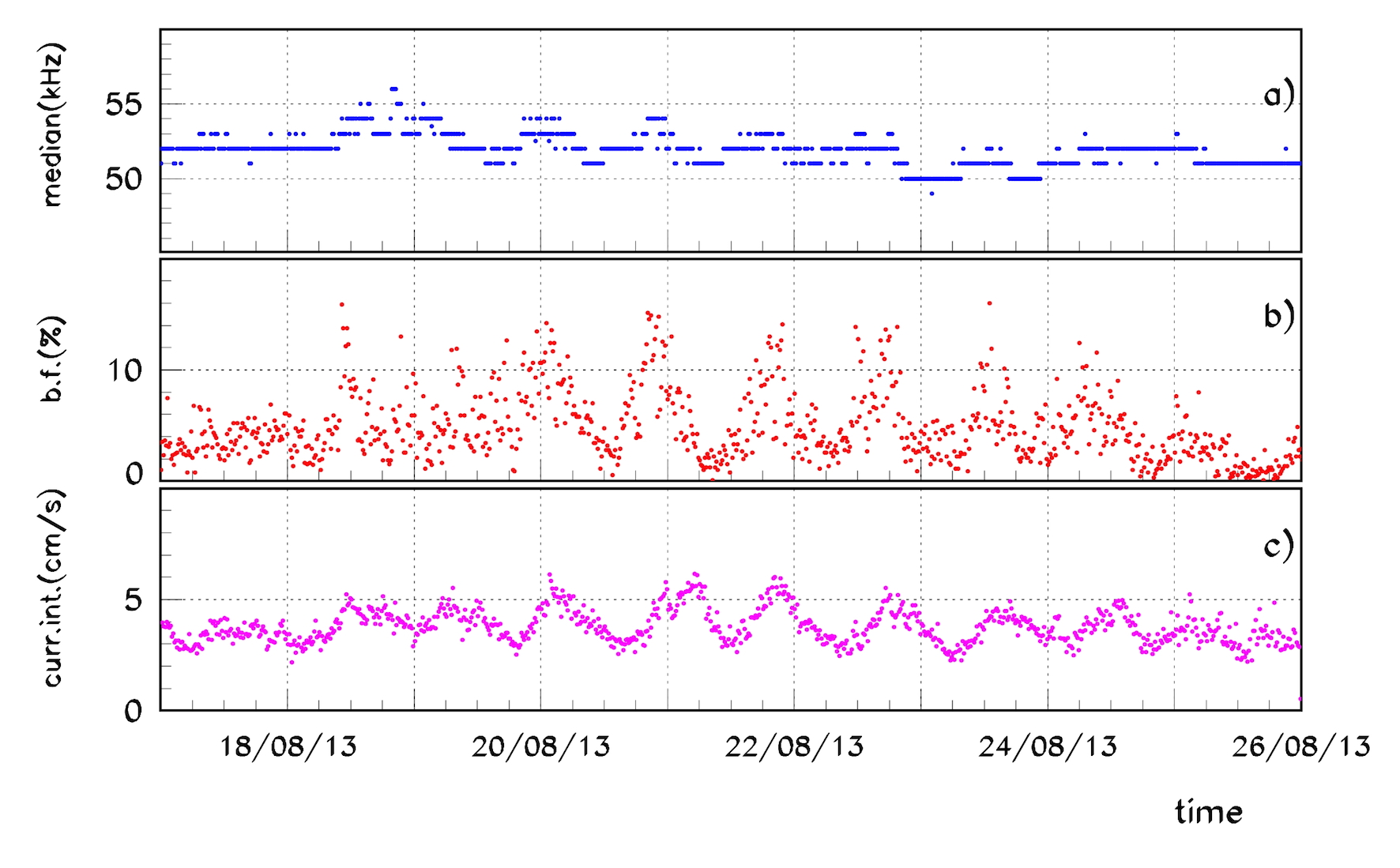}%
\end{minipage}
\caption{a) baseline rates over a 15 minutes period; b) burst fraction (b.f.), defined as the fraction of time in the 15 minute window with rate exceeding 100 kHz; c) modulus of the water current velocity measured with the DCS.}
\label{bussolepulse}
\end{figure}

The time behaviour of the small bioluminescence activity is shown in Fig.~\ref{bussolepulse} for a 15 days time series. There is evidence of periodic behaviour due to the presence of inertial currents which at the tower latitudes have a period of 20.21 h. Further analyses of bioluminescence correlation
with sea currents are in progress.

\section{Conclusions and perspectives}
Deployed in March 2013 in the abyssal site of Capo Passero at the depth of 3500 m, the NEMO Phase 2 tower continuously took data until August 2014, validating the technical solutions proposed for the construction of an underwater Cherenkov neutrino detector.
The data analyses have confirmed the optimal environmental characteristics of the KM3NeT-It deep-sea site.

This site will host the Italian node of the KM3NeT underwater neutrino telescope.
This detector will comprise eight tower-like detection units, built with a similar technology as the one used for the NEMO Phase-2 prototype described here but with 14 floors each one holding six optical modules, and 24 string-like detection units, using a new technology developed by the European KM3NeT collaboration.

\end{document}